\documentclass[prd,amsmath,twocolumn,floatfix,showpacs,preprintnumbers,superscriptaddress]{revtex4}
\usepackage{epsfig}
\usepackage{graphicx}

\newcommand{\rf}[4]{{#1} {\bf #2}, #3 (#4)}

\newcommand{\cm}{Commun.\ Math.\ Phys.}

\newcommand{\al}{\alpha}
\newcommand{\mr}{\mathrm}

\newcommand{\tr}{\text{Tr}}

\newcommand{\be}{\begin{equation}}
\newcommand{\ee}{\end{equation}}
\newcommand{\slh}{\!\!\!\slash}

\newcommand{\ZI}{Z^{(\rm{I})}(p)}
\newcommand{\ZR}{Z^{(\rm{R})}(p)}

\begin{document}

\preprint{ADP-09-001/T679}

\title{
Stout-link smearing in lattice fermion actions
}

\author{J.~B.~Zhang}
\affiliation{ZIMP and Department of Physics, Zhejiang University,
    Hangzhou, 310027, P. R. China}
\affiliation{Special Research Center for the Subatomic Structure of Matter
    (CSSM) and Department of Physics,\\ University of Adelaide 5005,
    Australia}
\author{Peter~J.~Moran}
\affiliation{Special Research Center for the Subatomic Structure of Matter
    (CSSM) and Department of Physics,\\ University of Adelaide 5005,
    Australia}
\author{Patrick~O.~Bowman}
\affiliation{Centre for Theoretical Chemistry and Physics and
  Institute of Natural Sciences, Massey University (Albany), 
  Private Bag 102904, North Shore City 0745, New Zealand}
\author{Derek~B.~Leinweber}
\author{Anthony~G.~Williams}
\affiliation{Special Research Center for the Subatomic Structure of Matter
    (CSSM) and Department of Physics,\\ University of Adelaide 5005,
    Australia}

\date{\today}


\begin{abstract}
The properties of the momentum space quark propagator in Landau gauge
are studied for the overlap quark action in quenched lattice
QCD. Numerical calculations are performed over four ensembles of gauge
configurations, where three are smeared using either 1, 3, or 6 sweeps
of stout-link smearing. We calculate the non-perturbative wave
function renormalization function $Z(p)$ and the non-perturbative mass
function $M(p)$ for a variety of bare quark masses. We find that the
wave-function renormalization function is slightly sensitive to the
number of stout-link smearing sweeps. For the mass function we find
the effect of the stout-link smearing algorithm to be small for
moderate to light bare quark masses. For a heavy bare quark mass we
find a strong dependence on the number of smearing sweeps.
\end{abstract}


\maketitle


The quark propagator is one of the fundamental components of Quantum
Chromodynamics (QCD). Although it is not a physical observable, many
physical quantities are related to it. By studying the
momentum-dependent quark mass function in the infrared region we can
gain valuable insights into the mechanisms of dynamical chiral
symmetry breaking and the associated dynamical generation of mass.
At high momenta, one can also use the quark propagator to extract the
running quark mass~\cite{Bowman:2004xi}.

Lattice QCD techniques provide an avenue for the non-pertubative study
of the quark propagator. There have been several lattice studies of
the momentum space quark
propagator~\cite{soni1,soni2,jon1,jon2,Bow02a,Bow02b,blum01,overlgp,overlgp2,qpscaling}
using different fermion actions. Finite volume effects and
discretization errors have also been extensively explored in lattice
Landau gauge~\cite{overlgp2,qpscaling}.

The overlap fermion formalism~\cite{neuberger0,neuberger2} realizes
an exact chiral symmetry on the lattice and is automatically ${\cal
O}(a)$ improved. There are many salient features of overlap fermions,
which include no additive renormalizations to the quark masses, an
index theorem linking the number of zero-modes of the Dirac operator
to the topological charge $Q$, and evading the so called ``no-go
theorem'' etc., however they are rather computationally demanding. There
are many suggestions on how to reduce the computational cost. One
such proposal is the use of a more elaborate kernel, together with a
fattening of the gauge 
links~\cite{Bietenholz:2002ks,DeGrand:2000tf,Kamleh:2001ff}.

The idea of any UV-filtered fermion action~\cite{degrand1, degrand2,
  Durr1, Durr2, Durr3} is that one will carry out the calculation on a
smoothed copy of the actual gauge field and evaluate the Dirac
operator on that background.  This yields a new fermion action which
differs from the old one by terms which are both simultaneously
\emph{ultralocal} and \emph{irrelevant}.  The term ``UV-filtered''
indicates that such an action is less sensitive to the UV fluctuations
of the gauge background. Sometimes, one also speaks of ``fat-link''
actions.

There is a great amount of freedom available when generating a
smoothed copy of some gauge field. One needs to decide on the
smoothing recipe (APE~\cite{Albanese:1987ds},
HYP~\cite{Hasenfratz:2001hp}, stout-link~\cite{Morningstar:2003gk},
etc.), on the parameter
($\al^\mr{APE},\al_{1,2,3}^\mr{HYP}$,$\al^\mr{SL}\equiv\rho$) and on
the number of iterations, $n^\mr{iter}$. In any case, with fixed
$(\al,n^\mr{iter})$ the filtered ``fat-link'' action is in the
\emph{same universality class} as the usual ``thin-link''
version~\cite{Durr2}.  Unfortunately, if any smoothing process is
over-applied, some important properties of the theory are lost.
Therefore, one needs to find a balance between the smoothing
procedure, which will accelerate convergence of the quark operator
inversion and improve the
localization properties, at the danger of losing important
physics. Recently, Stephan Durr and collaborators~\cite{Durr1, Durr2,
  Durr3} applied 1-3 sweeps stout-link
smearing~\cite{Morningstar:2003gk} to the lattice gauge configurations
and analysed how this affected various physical quantities. They claim
that it is safe to use 1-3 sweeps of standard stout-link smearing on
the gauge configurations.
More recently, 6 sweeps of stout-link smearing was used in the
\emph{Science} article exploring the hadron mass
spectrum~\cite{durr_science}.

In this paper, we investigate the momentum space quark propagator on
quenched gauge configurations. We utilise both the original lattice
configurations and also the configurations which are produced by one,
three, and six sweeps of standard stout-link smearing respectively. We
compare results across all four cases, in order to explore the effect
of smearing on the quark propagator with different quark masses,
different lattice momenta, etc.

The massive overlap operator can be written as~\cite{edwards2}
\begin{eqnarray}
D(\mu) = \frac{1}{2}\left[1+\mu+(1-\mu)\gamma_5 \epsilon(H_w)
\right] \, , \label{D_mu_eqn}
\end{eqnarray}
where $H_w(x,y)=\gamma_5 D_w(x,y)$ is the Hermitian Wilson-Dirac
operator,  $\epsilon(H_w)$ = $H_w/\sqrt{H_w^2}$ is the matrix sign
function, and the dimensionless quark mass parameter $\mu$ is
\begin{equation}
\mu \equiv \frac{m^0}{2m_w} \, , \label{mu_defn}
\end{equation}
where $m^0$ is the bare quark mass and $m_w$ is the Wilson quark
mass which, in the free case, must lie in the range $0 < m_w < 2$.

The bare quark propagator in coordinate space is given by
\begin{equation}
S^{\rm bare}(m^0)\equiv \tilde{D}_c^{-1}(\mu) \, ,
\label{overlap_propagator}
\end{equation}
where
\begin{equation*}
\tilde{D}_c^{-1}(\mu) \equiv \frac{1}{2m_w} \tilde{D}^{-1}(\mu)
\hspace{0.5cm}{\rm{and}}
\end{equation*}
\be
\tilde{D}^{-1}(\mu) \equiv
\frac{1}{1-\mu}\left[{D}^{-1}(\mu)-1\right] \, . \label{D_mu}
\end{equation}

When all interactions are turned off, the inverse bare lattice
quark propagator reduces to the tree-level version, and in momentum
space is given by
\begin{equation}
(S^{(0)})^{-1}(p)\equiv
{i\left(\sum_{\mu}C_{\mu}^{(0)}(p)\gamma_{\mu}\right)+B^{(0)}(p)}\,,
\label{treeinvpro}
\end{equation}
where $p$ is lattice momentum. One can calculate $S^{(0)}(p)$
directly by setting all links to unity in coordinate space,
doing the matrix inversion, and then taking its Fourier transform.
It is then possible to identify the appropriate kinematic lattice
momentum $q$  directly from the definition
\begin{equation}
q_\mu\equiv C_{\mu}^{(0)}(p). \label{latmomt}
\end{equation}
The form of $q_\mu(p_\mu)$ is shown and its analytic form given in
Ref.~\cite{overlgp}.  Having identified the appropriate kinematical
lattice momentum $q$, we can now define the bare lattice propagator
as
\begin{equation}
S^{\rm bare}(p) \equiv \frac{Z(p)}{i{q\slh}+M(p)}.
\end{equation}
This ensures that the free lattice propagator is identical to the
free continuum propagator.  Due to asymptotic freedom the lattice
propagator will also take the continuum form at large momenta.  In
the gauge sector, this type of analysis dramatically improves the
gluon propagator~\cite{Lei99,Bon00,Bon01}.

The two Lorentz invariants can then be obtained via
\begin{gather}
Z^{-1}(p) = \frac{1}{12iq^2} \tr \{q\slh S^{-1}(p) \} \,, \\
M(p) = \frac{Z(p)}{12} \tr \{ S^{-1}(p) \} \,.
\end{gather}
Here $Z(p)$ is the wave-function renormalization function and $M(p)$
is the mass function. The above equations imply that $Z(p)$ is
directly dependent on our choice of momentum $q$, whilst $M(p)$ is
not. 

Standard stout-link smearing, using an isotropic smearing parameter
$\rho_{\rm sm}$, involves a simultaneous update of all links on the
lattice. Each link is replaced by a smeared link
$\tilde{U}_\mu(x)$~\cite{Morningstar:2003gk}
\begin{equation}
  \tilde{U}_\mu(x) = \mathrm{exp}(i Q_\mu(x) ) \, U_\mu(x) \,,
  \label{eqn:stoutlinksmearedlink}
\end{equation}
where
\begin{align}
  Q_\mu(x) & = \frac{i}{2}(\Omega_\mu^\dagger(x) - \Omega_\mu(x)) \notag \\
  & \quad - \frac{i}{6}\tr(\Omega_\mu^\dagger(x) - \Omega_\mu(x)) \,,
\end{align}
with
\begin{equation}
  \Omega_\mu(x) = \rho_{\rm sm}\,\sum\{1\times1{\rm\ loops\ involving\ }U_\mu(x) \} \,.
\end{equation}

We work on $16^3 \times 32$ lattices, with gauge configurations
created using a tadpole improved, plaquette plus rectangle
(L\"{u}scher-Weisz~\cite{Lus85}) gauge action through the
pseudo-heat-bath algorithm. The lattice spacing, $a$ = $0.093$ fm, is
determined from the static quark potential with a string tension of
$\sqrt{\sigma}$ = $440$~MeV~\cite{zanotti}. The number of
configurations to be used for each ensemble in this study is $50$.
The first smeared ensemble is created by applying one sweep of
stout-link smearing to the original configurations with a smearing
parameter of $\rho$ = $0.10$. The second smeared ensemble is created
using three sweeps of stout-smearing with the same value of $\alpha$.
We work in an $\mathcal{O}(a^2)$-improved Landau gauge, and fix the
gauge using a Conjugate Gradient Fourier Acceleration~\cite{cm}
algorithm with an accuracy of
$\theta\equiv\sum\left|\partial_{\mu}A_{\mu}(x)\right|^{2}<10^{-12}$.
The improved gauge-fixing scheme was used to minimize gauge-fixing
discretization errors~\cite{bowman2}.

Our numerical calculation begins with an evaluation of the inverse
of $D(\mu)$ on the unfixed gauge configurations, where $D(\mu)$ is
defined in Eq.~(\ref{D_mu_eqn}). We then calculate the quark
propagator of Eq.~(\ref{overlap_propagator}) for each configuration
and rotate it to the Landau gauge by using the corresponding gauge
transformation matrices \{$G_i(x)$\}. We then take the ensemble
average to obtain $S^{\rm bare}(x,y)$. The discrete Fourier
transformation is then 
applied to $S^{\rm bare}(x,y)$ and the momentum-space bare quark
propagator, $S^{\rm bare}(p)$, is finally obtained.

We use the mean-field improved Wilson action in the overlap fermion
kernel. The value $\kappa$ = $0.19163$ is used in the Wilson action,
which provides $m_w a$ = $1.391$ for the Wilson regulator mass in
the interacting case~\cite{overlgp}. We calculate the overlap quark
propagator for $15$ bare quark masses on each ensemble by using a
shifted Conjugate Gradient solver. The bare quark mass $m^0$ is
defined by Eq.~(\ref{mu_defn}). In the calculation, we choose the
mass parameter $\mu =$ $0.009$, $0.010$, $0.012$, $0.014$, $0.016$,
$0.018$, $0.021$, $0.024$, $0.030$, $0.036$, $0.045$, $0.060$,
$0.075$, $0.090$, and $0.105$. This choice of $\mu$ corresponds to
bare quark masses, in physical units, of $m^0 = $ $53$, $59$, $71$,
$82$, $94$, $106$, $124$, $142$, $177$, $212$, $266$, $354$, $442$,
$531$, and $620$~MeV respectively.

The partial results for the mass function $M(p)$ and the
wave-function renormalization function $Z^{(\rm{R})}(p)\equiv
Z(\zeta;p)$ on a $16^3\times 32$ lattice without any smearing in
Landau gauge were reported in Ref.~\cite{qpscaling}. Here we 
focus on a comparison of the behavior of the overlap fermion
propagator when using different numbers of stout-link smearing
sweeps. All data is cylinder cut~\cite{Lei99}. Statistical
uncertainties are estimated via a second-order, single-elimination
jackknife. 

In a standard lattice simulation, one begins by tuning the value of
the input bare quark mass $m^0$ to give the desired renormalized quark
mass, which is usually realized through the calculation of a physical
observable.  However, smearing a lattice configuration filters out the
ultraviolet physics and the renormalization of the mass will be
different. To some extent, the effect is similar to that of an
increase in the lattice spacing $a$.  After smearing, the same input
$m^0$ will therefore give a different renormalized quark mass.  The
input bare quark mass must then be re-tuned in order to reproduce the
same physical behavior as on the unsmeared configuration.

We wish to directly study how the quark propagator $S(p)$ is affected
by smearing, through a calculation of the mass $M(p)$ and
wave-renormalization $Z(p)$ functions.  In order to replicate the
re-tuning procedure described above, we begin by first calculating
$M(p)$ and $Z(p)$ for all values of $m^0$ listed previously, over all
four types of configurations.  We then select a value of the bare
quark mass $m^0$ to investigate, and force the mass functions $M(p)$
to agree at a given reference momentum, $\zeta$.  This is
achieved by interpolating $M(p)$, for the smeared configurations,
between neighboring values of the bare quark masses, in order to
determine the required effective bare quark mass.  Any reasonable
choice of $\zeta$ should suffice.  By reasonable, we mean any point
out of the far infrared (IR) or ultraviolet (UV) momentum regions,
where lattice artifacts will spoil the results.

In comparing the renormalization function, we first interpolate $Z(p)$
to the effective bare quark mass, obtaining $\ZI$. We then
multiplicatively renormalize $\ZI$ to $\ZR \equiv Z(\zeta,p)$,
subject to $Z(\zeta,\zeta) = 1$.

\begin{figure*}
\centering
\begin{tabular}{c@{\hspace{0.05\textwidth}}c}
\includegraphics[angle=90,width=0.45\textwidth]{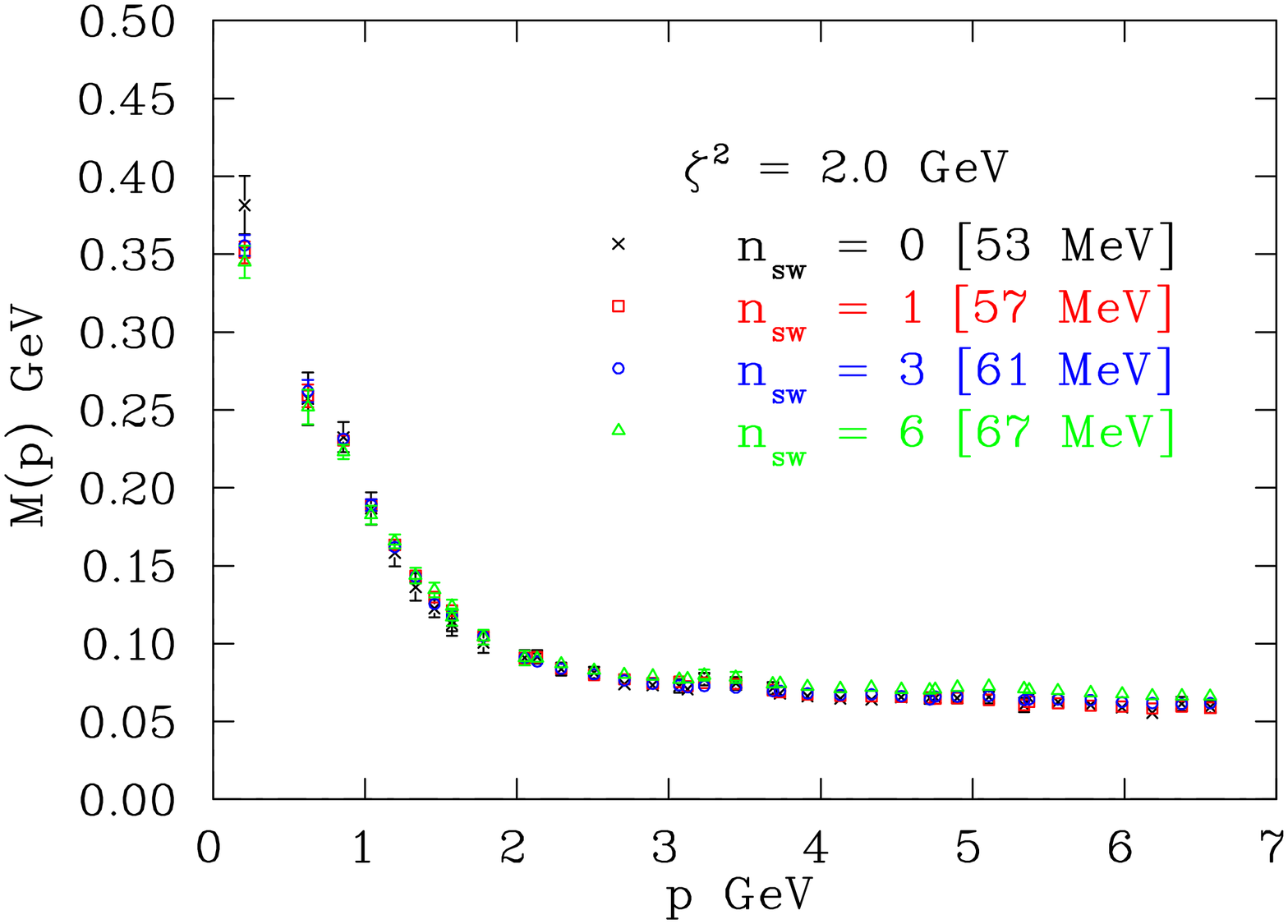} & 
\includegraphics[angle=90,width=0.45\textwidth]{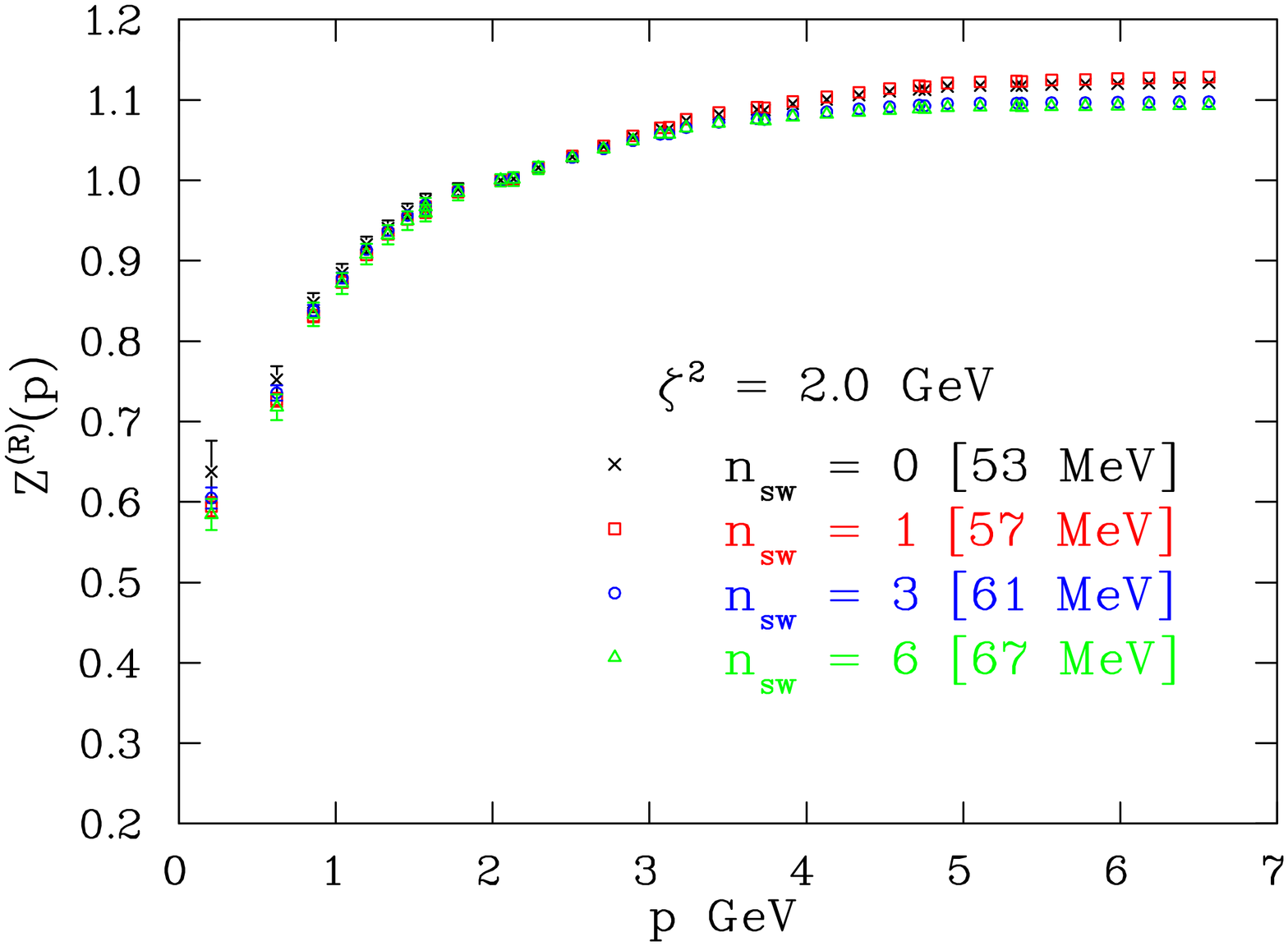} \\
\includegraphics[angle=90,width=0.45\textwidth]{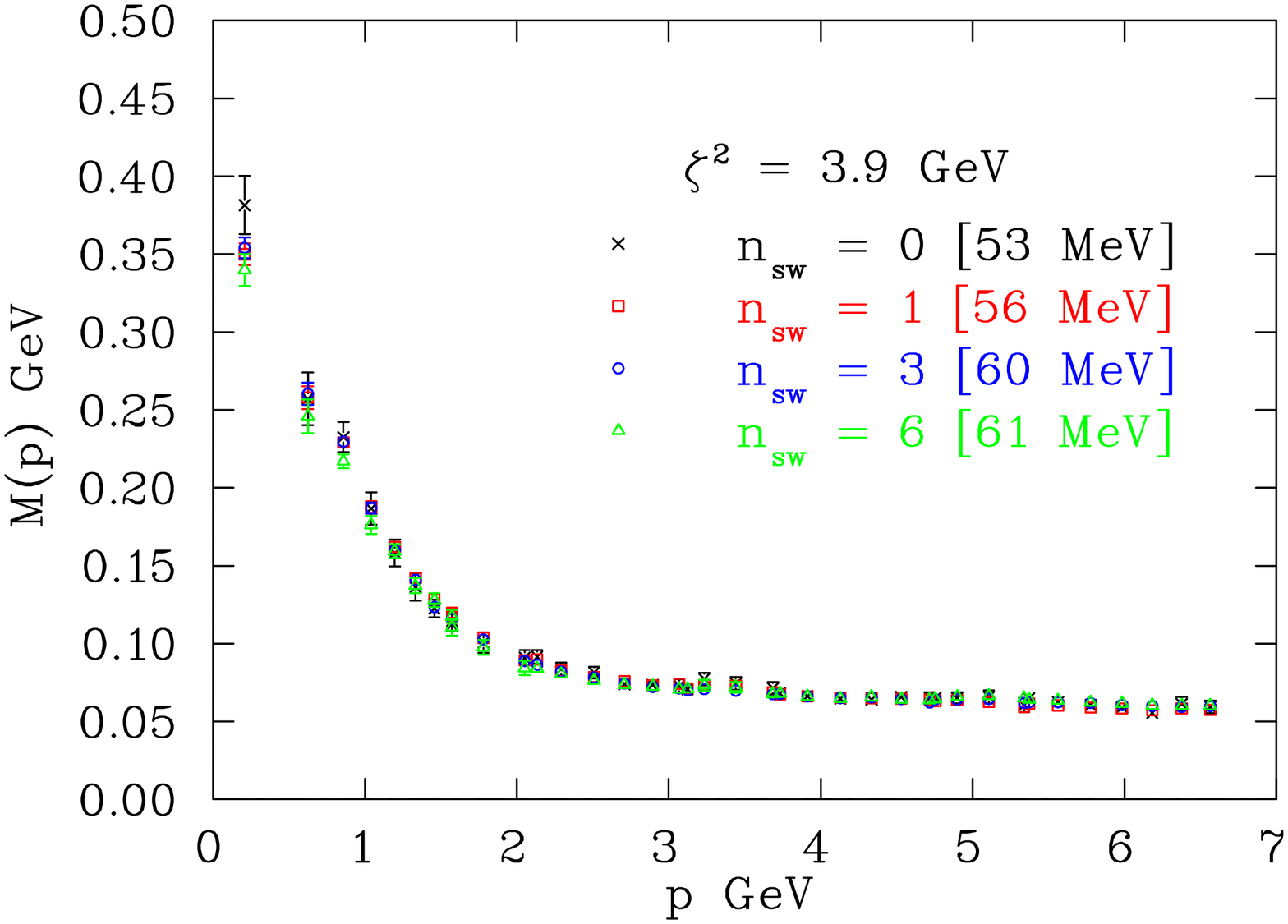} & 
\includegraphics[angle=90,width=0.45\textwidth]{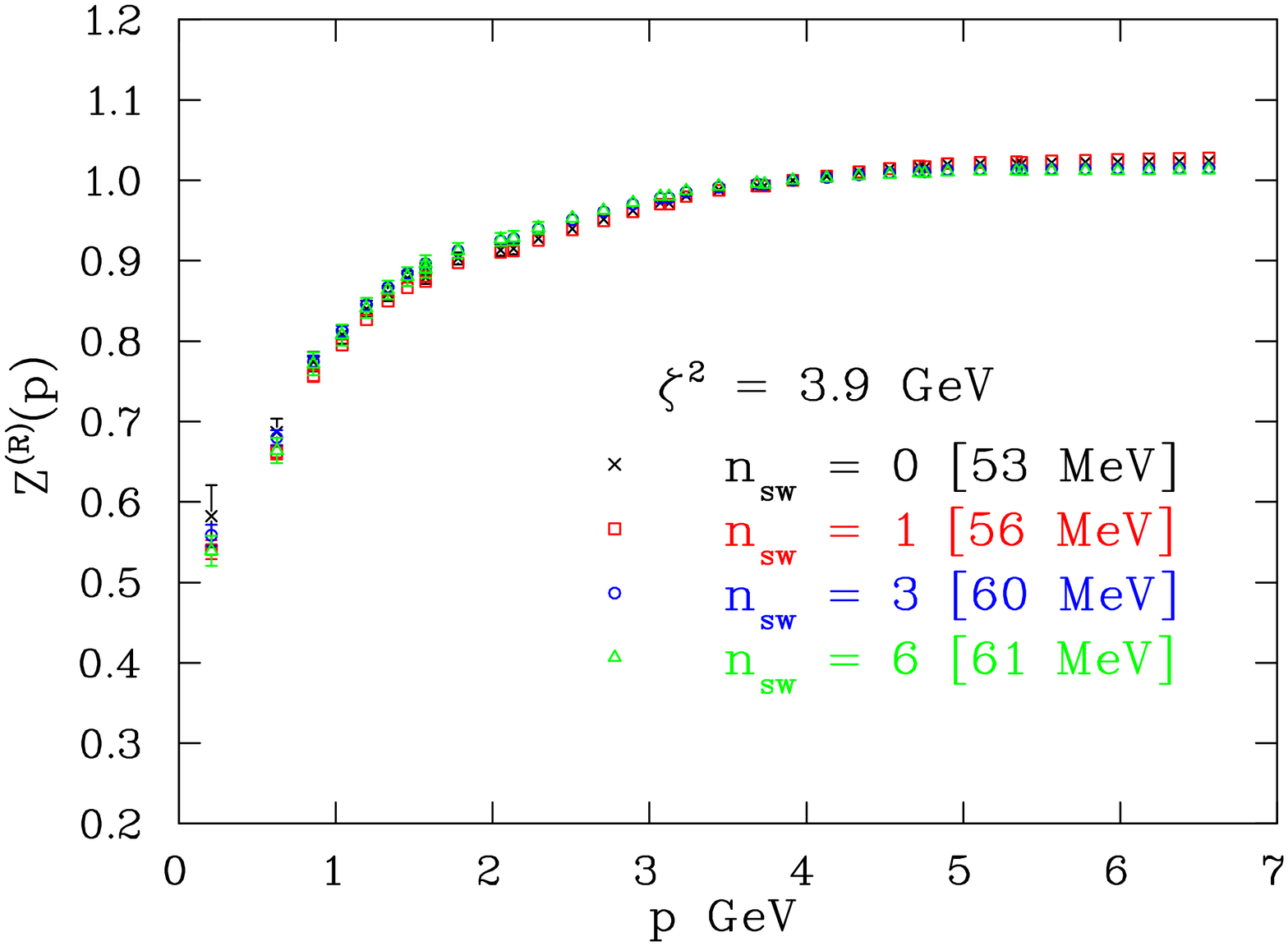} \\
\includegraphics[angle=90,width=0.45\textwidth]{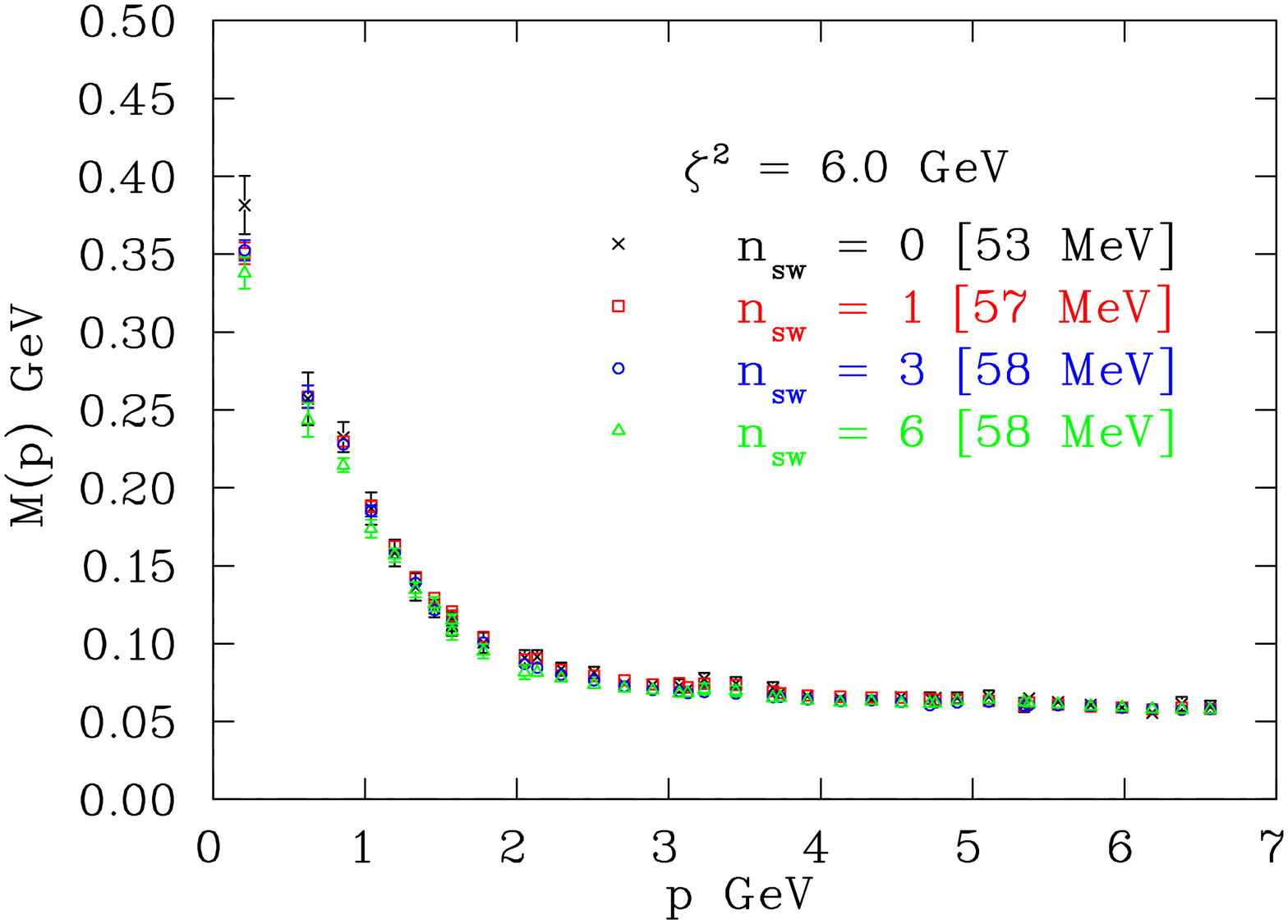} & 
\includegraphics[angle=90,width=0.45\textwidth]{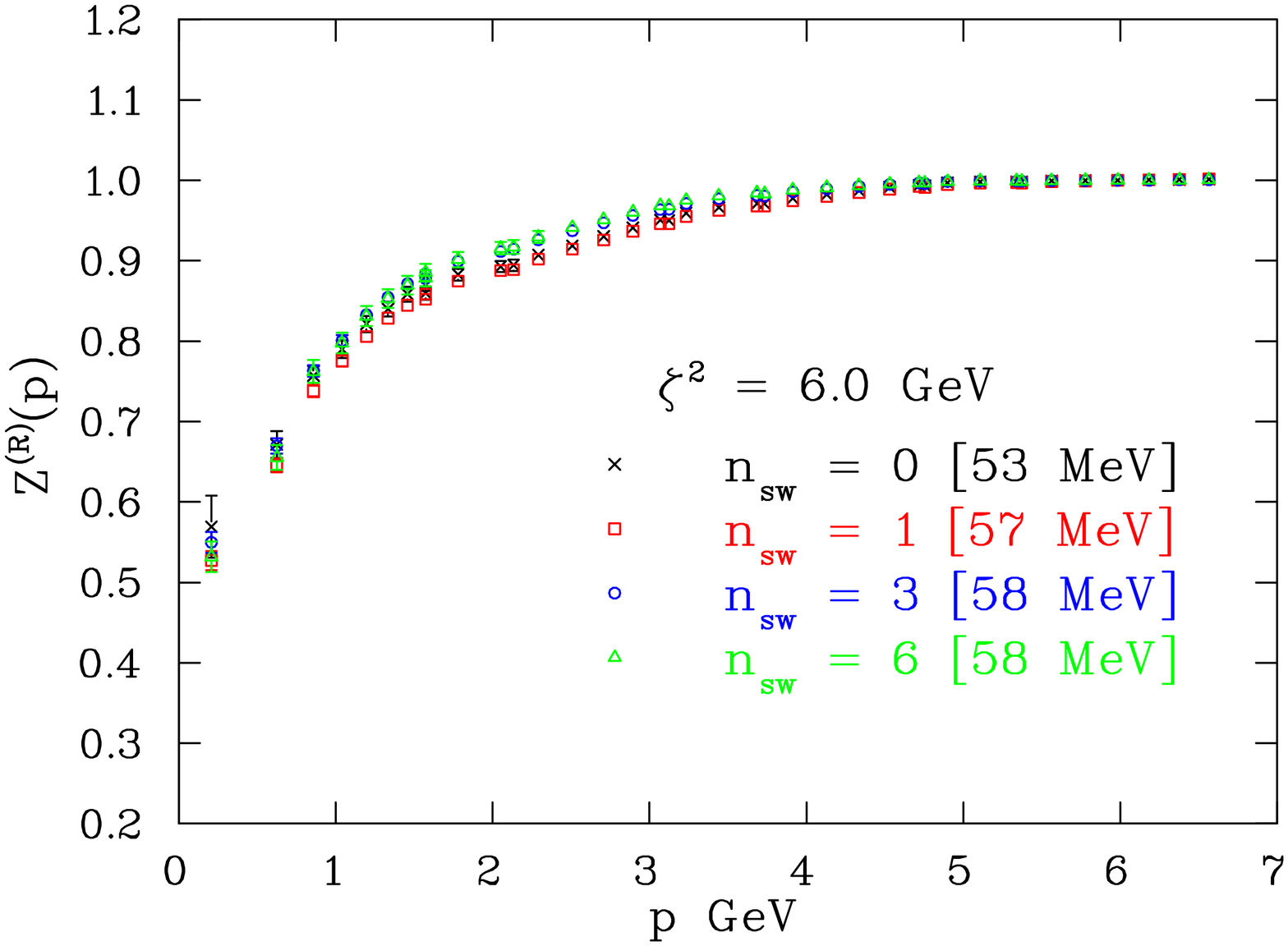}
\end{tabular}
\caption{The interpolated mass $M(p)$ and renormalization $\ZR$
  functions for the small bare quark mass, $m^0 = 53$~MeV, with 
  three choices of $\zeta$. The effective
  bare quark masses are given in square brackets. There is good
  agreement in $M(p)$ for all choices of $\zeta$ with up to six
  sweeps of stout-link smearing. A small splitting in the UV region of
  $\ZR$ is apparent after three sweeps of smearing. This leads to a
  disagreement in $\ZR$ for a large choice of $\zeta = 6.0$~GeV.}
\label{0090res}
\end{figure*}

We begin with a comparison of the functions $M(p)$ and
$Z^{(\rm{R})}(p)$ for a small bare quark mass, with three choices of
the reference momentum $\zeta = 2.0$, $3.9$ and $6.0$~GeV.  The
interpolated mass functions for the smallest bare quark mass $m^0 =
53$~MeV are given in Fig.~\ref{0090res}.  We note the significant
reduction in the statistical error, even after a single sweep of
smearing.  For all choices of $\zeta$, the mass functions display
strong agreement over all four levels of smearing, with the only
differences occurring in the most infrared points.  For the function
$Z^{(\rm{R})}(p)$ the effect of smearing is also subtle, however the
link smearing does introduce a minor splitting in the UV region.  This
splitting leads to small differences in the lower momentum regions of
$\ZR$ when we select $\zeta = 6.0$~GeV.

\begin{figure*}
\begin{tabular}{c@{\hspace{0.05\textwidth}}c}
\includegraphics[angle=90,width=0.45\textwidth]{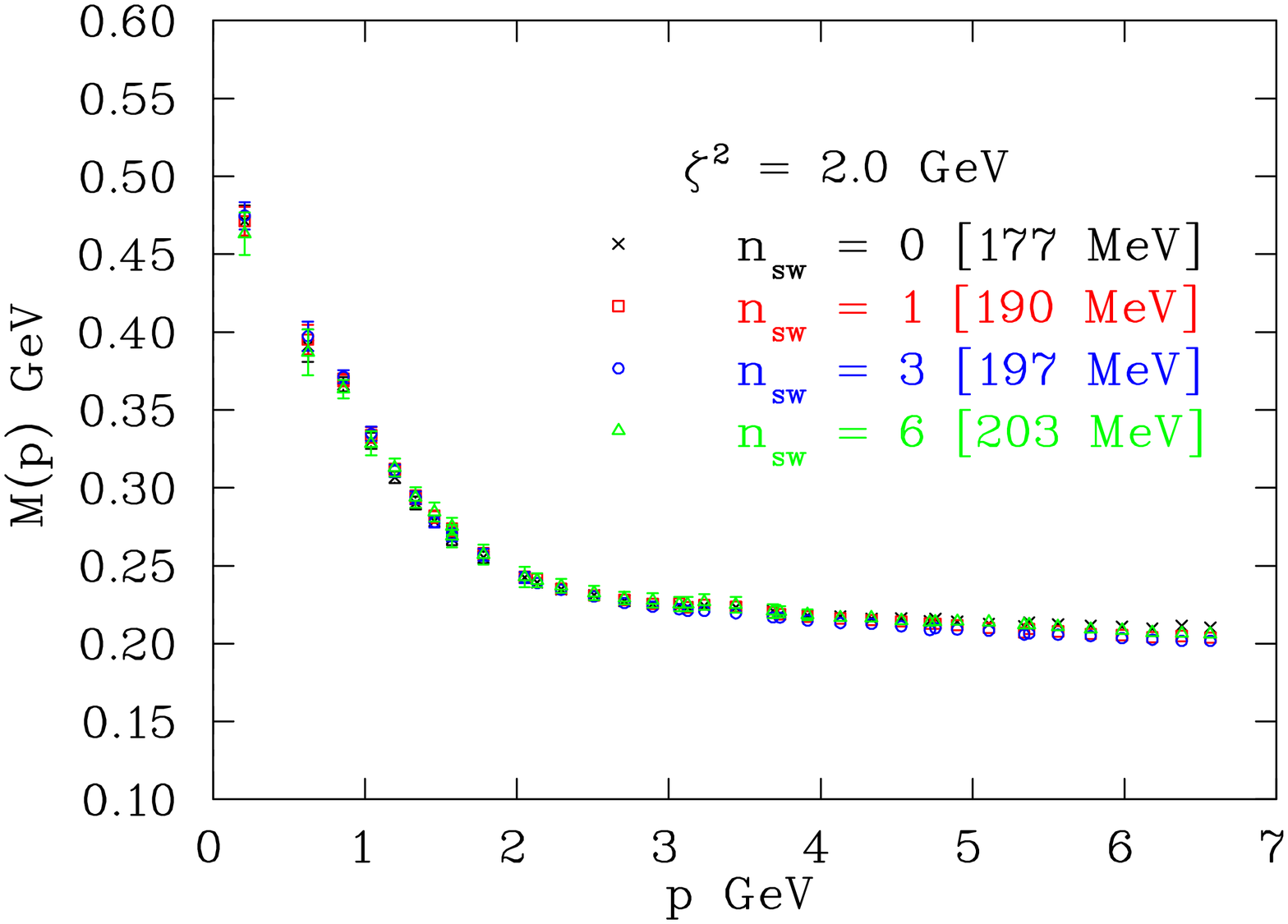} & 
\includegraphics[angle=90,width=0.45\textwidth]{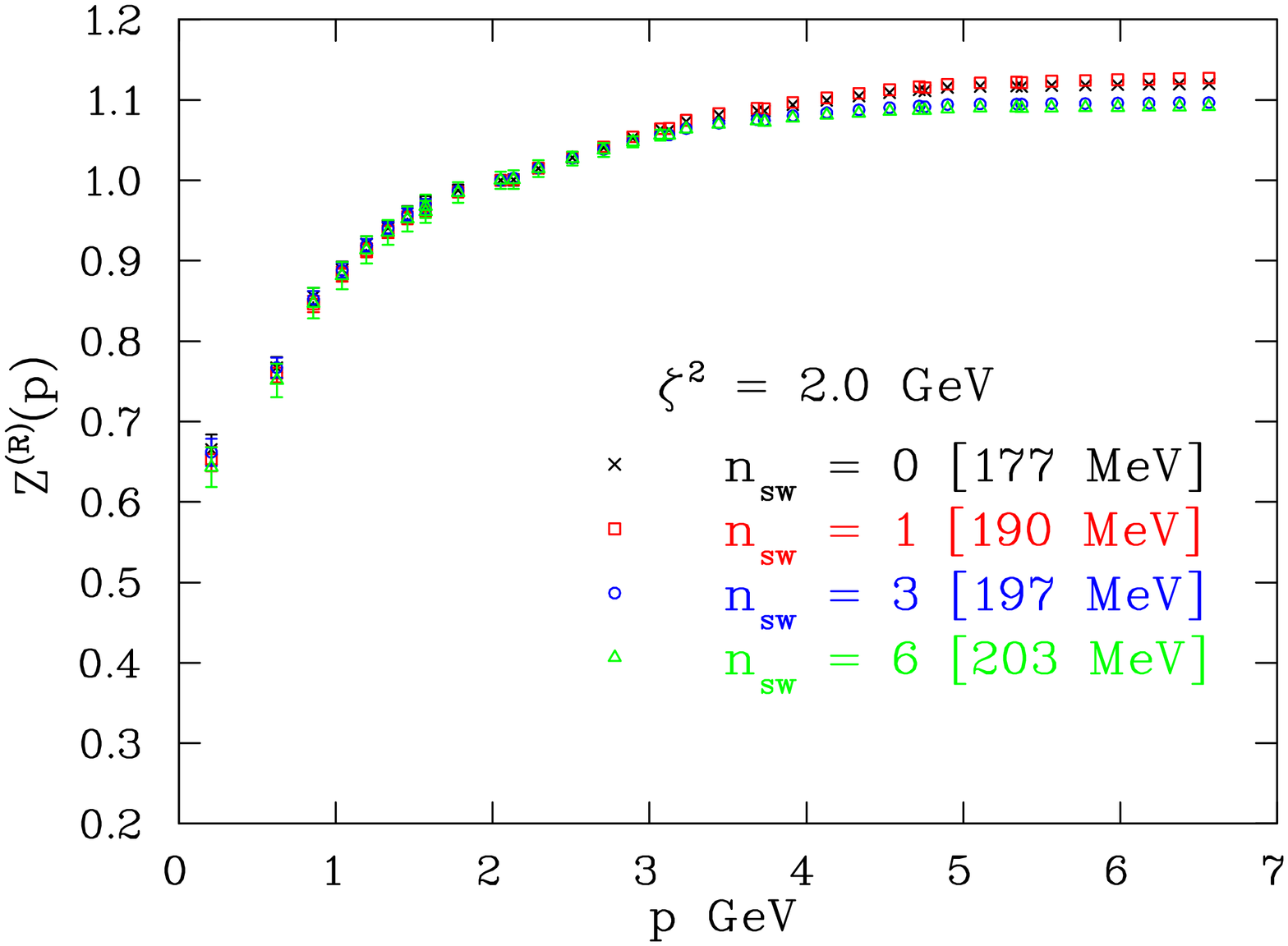} \\
\includegraphics[angle=90,width=0.45\textwidth]{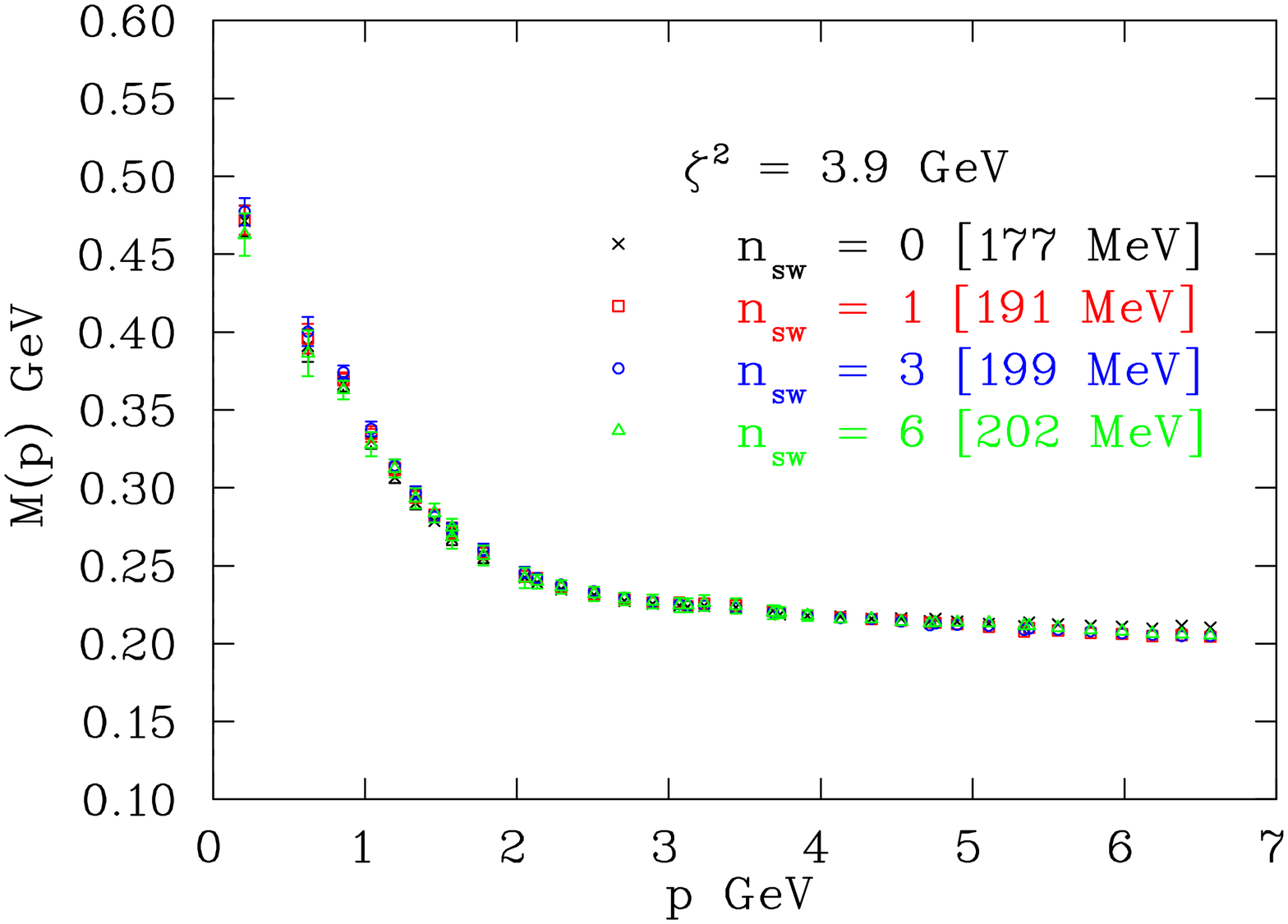} & 
\includegraphics[angle=90,width=0.45\textwidth]{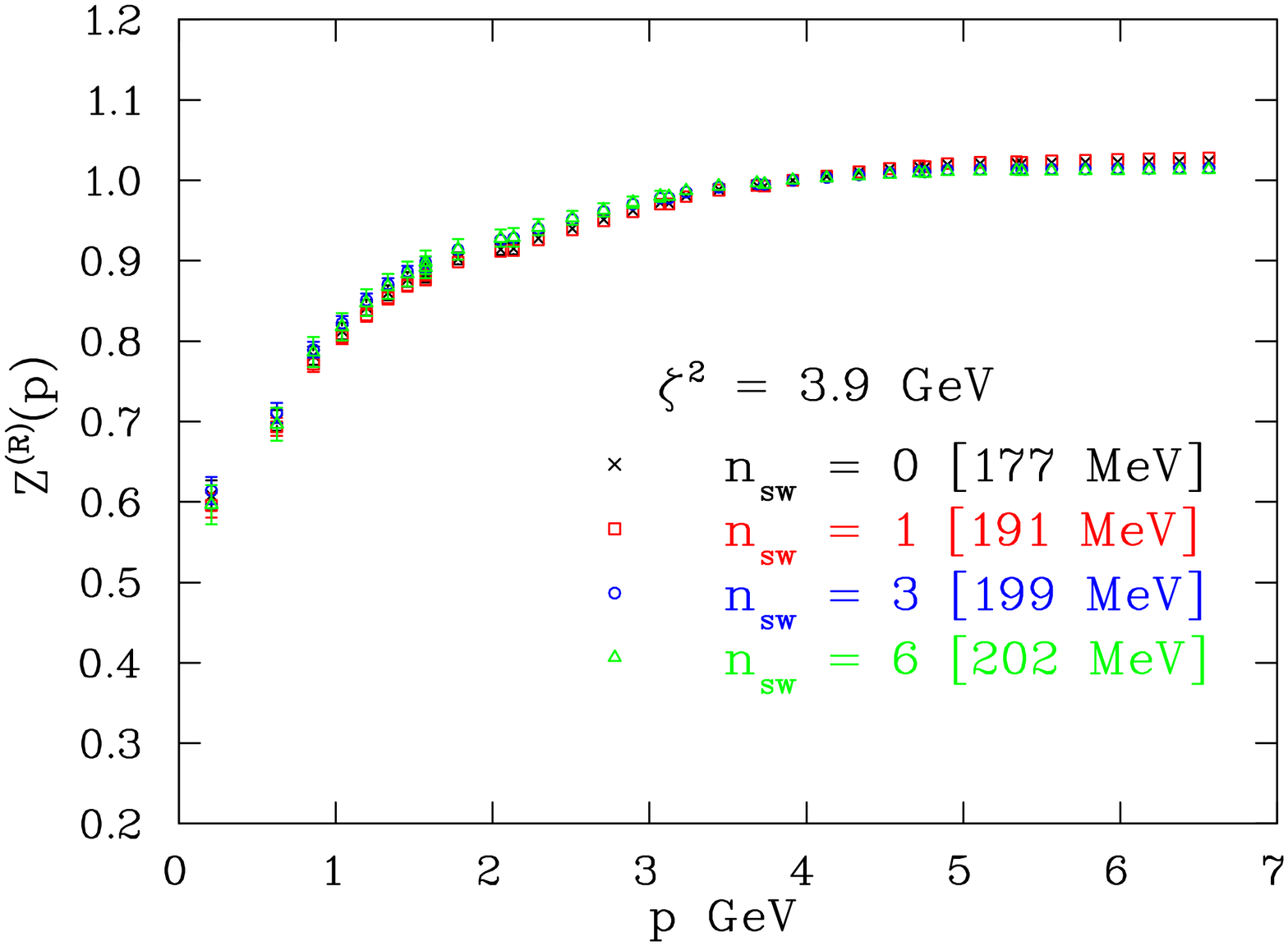} \\
\includegraphics[angle=90,width=0.45\textwidth]{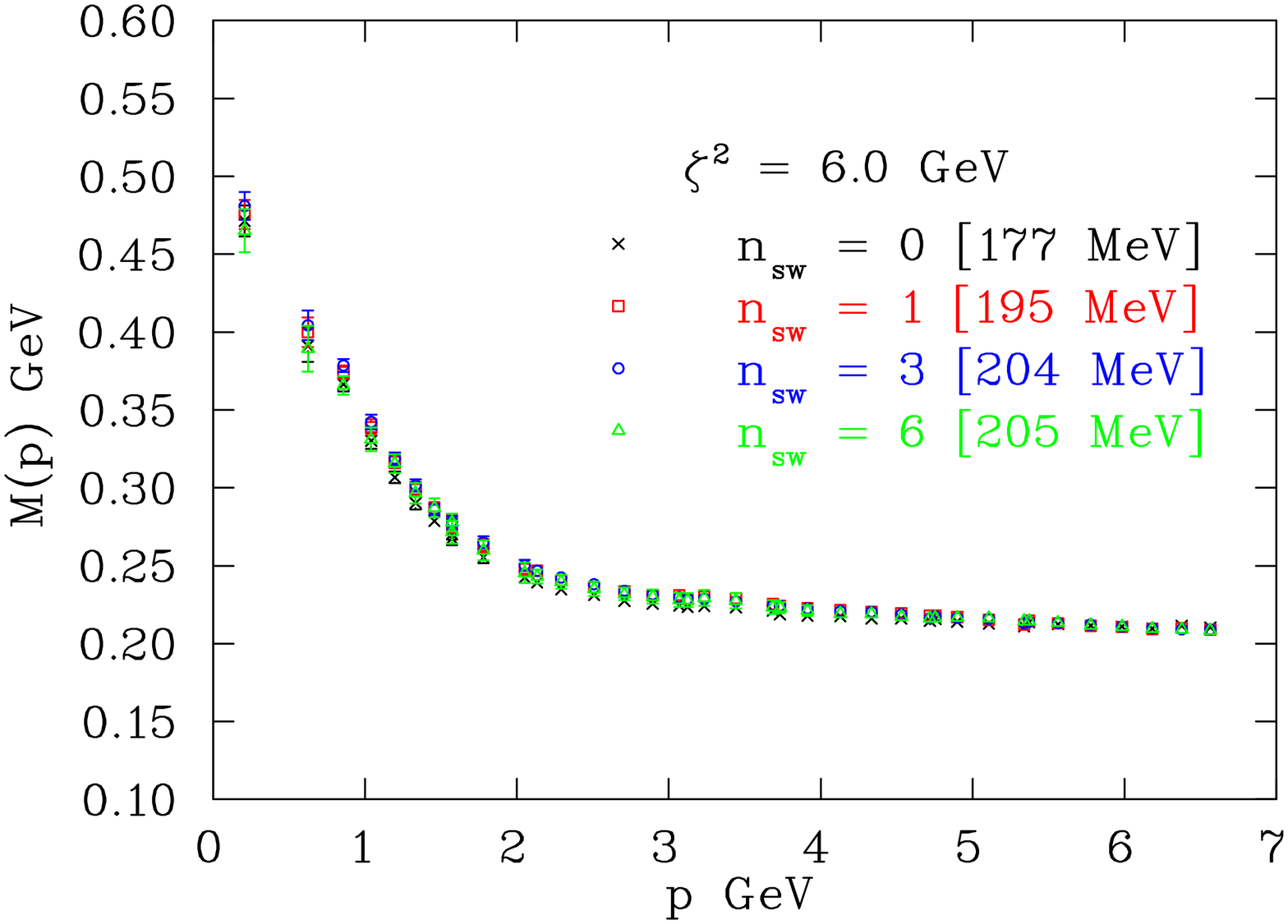} & 
\includegraphics[angle=90,width=0.45\textwidth]{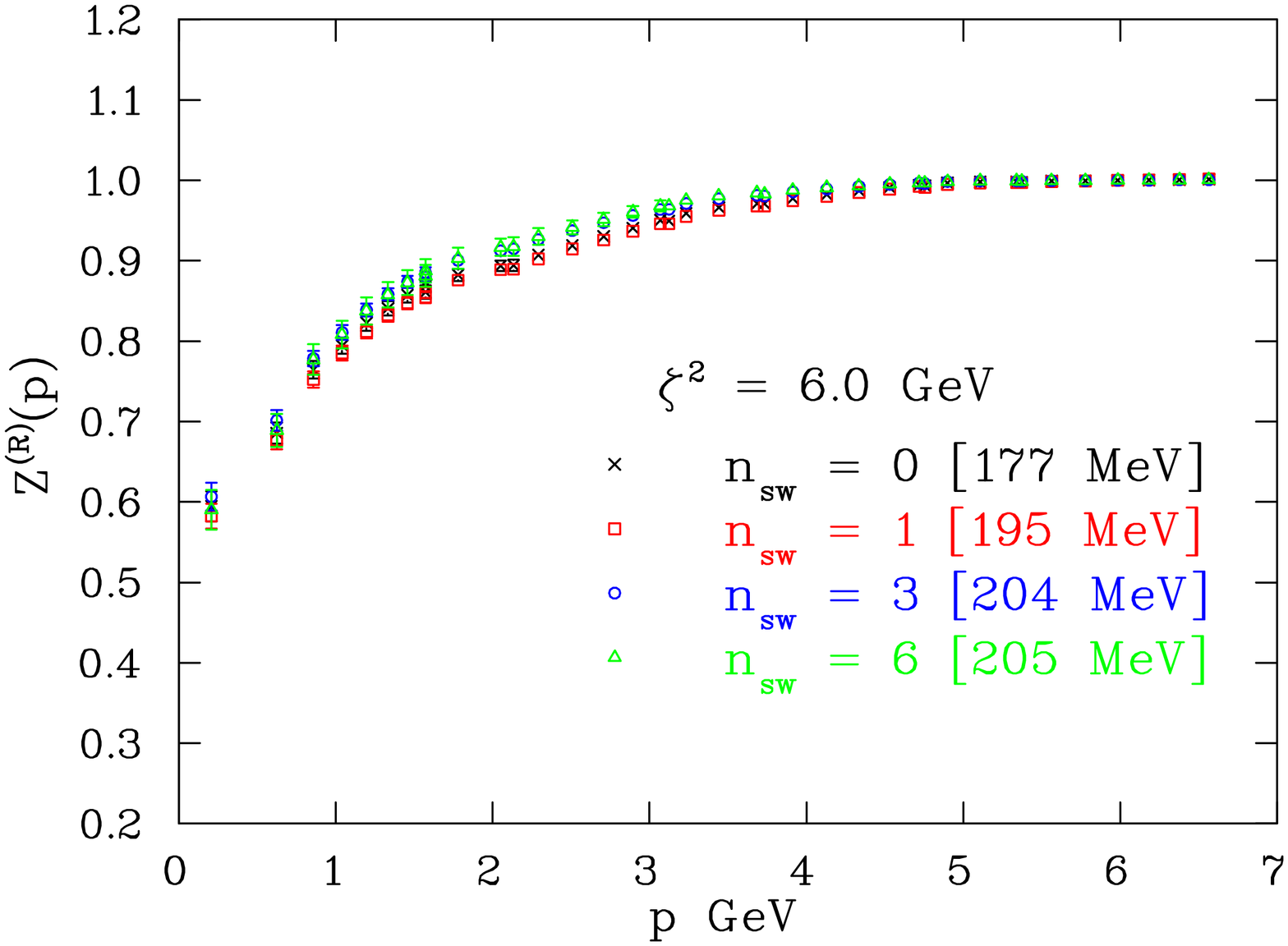}
\end{tabular}
\caption{The interpolated mass $M(p)$ and renormalization $\ZR$
  functions for the moderate bare quark mass, $m^0 = 177$~MeV, with
  the three choices of $\zeta$. The
  effective bare quark masses are given in square brackets. As with
  the small bare quark mass, the mass function displays good agreement
  for all choices of $\zeta$, and there is also a small splitting
  apparent in the UV region of $\ZR$. We note that the differences in
  $\ZR$ appear to be independent of the bare quark mass.}
\label{0300res}
\end{figure*}

Next we consider a moderate bare quark mass of $177$~MeV, for which
the functions $M(p)$ and $\ZR$ are shown in Fig.~\ref{0300res}.  As in
the case of a small bare quark mass, we find that the mass function
appears independent of the choice of reference momentum, however
the discrepancy at the most infrared point is no longer apparent.  The
renormalization function displays the same splitting in the UV region.
The effect of smearing on the quark propagator still appears to be
relatively minor at this value of $m^0$.

\begin{figure*}
\begin{tabular}{c@{\hspace{0.05\textwidth}}c}
\includegraphics[angle=90,width=0.45\textwidth]{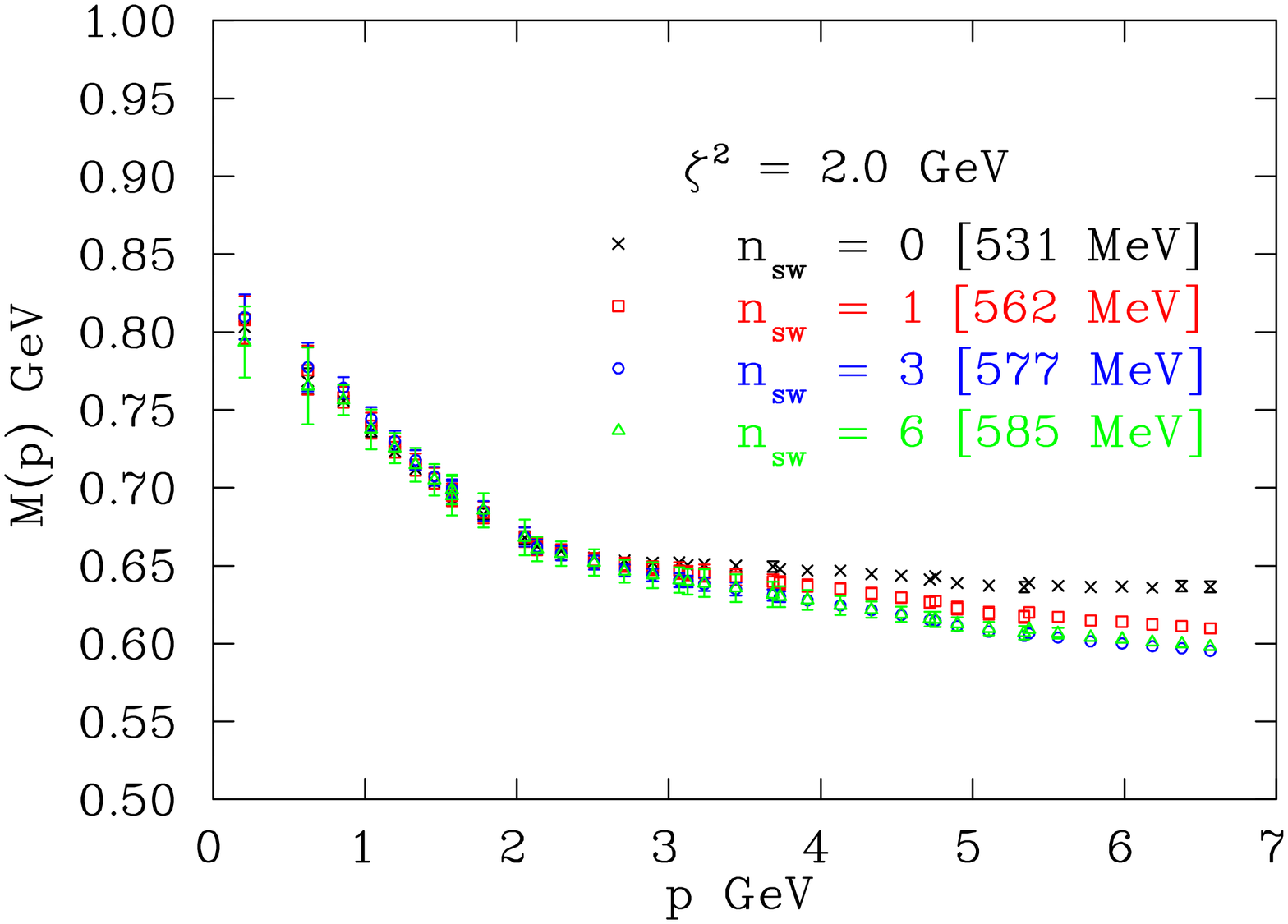} & 
\includegraphics[angle=90,width=0.45\textwidth]{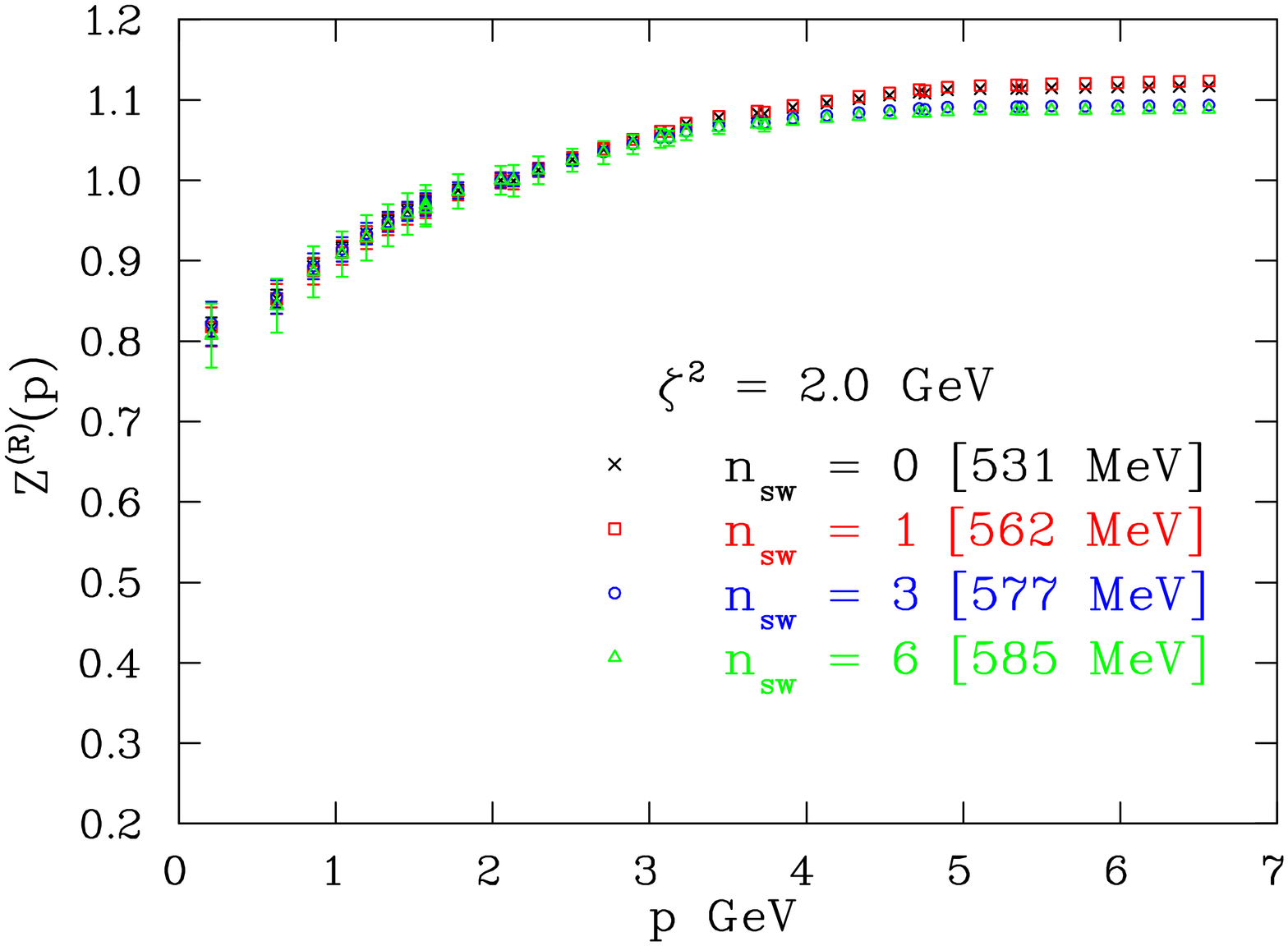} \\
\includegraphics[angle=90,width=0.45\textwidth]{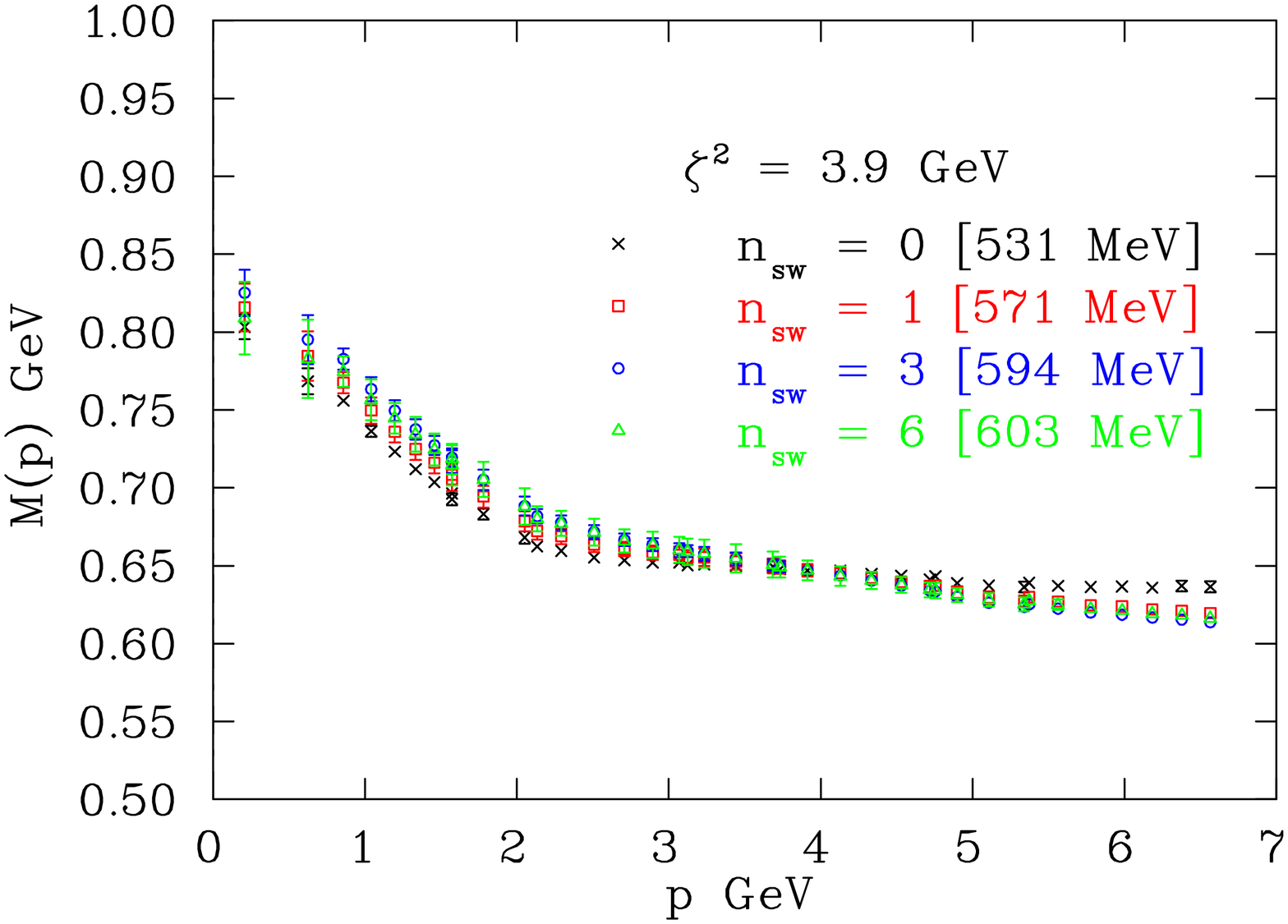} & 
\includegraphics[angle=90,width=0.45\textwidth]{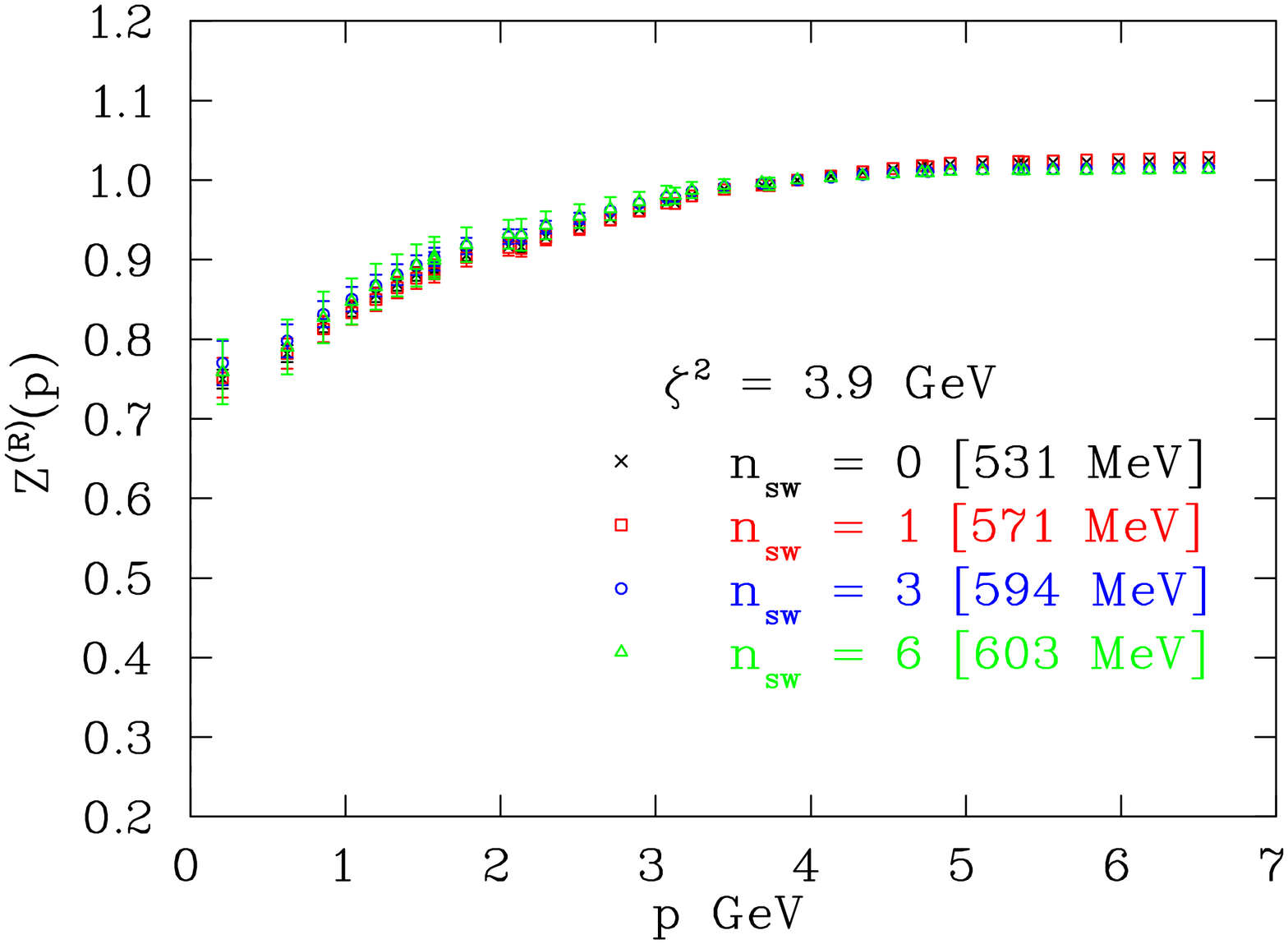} \\
\includegraphics[angle=90,width=0.45\textwidth]{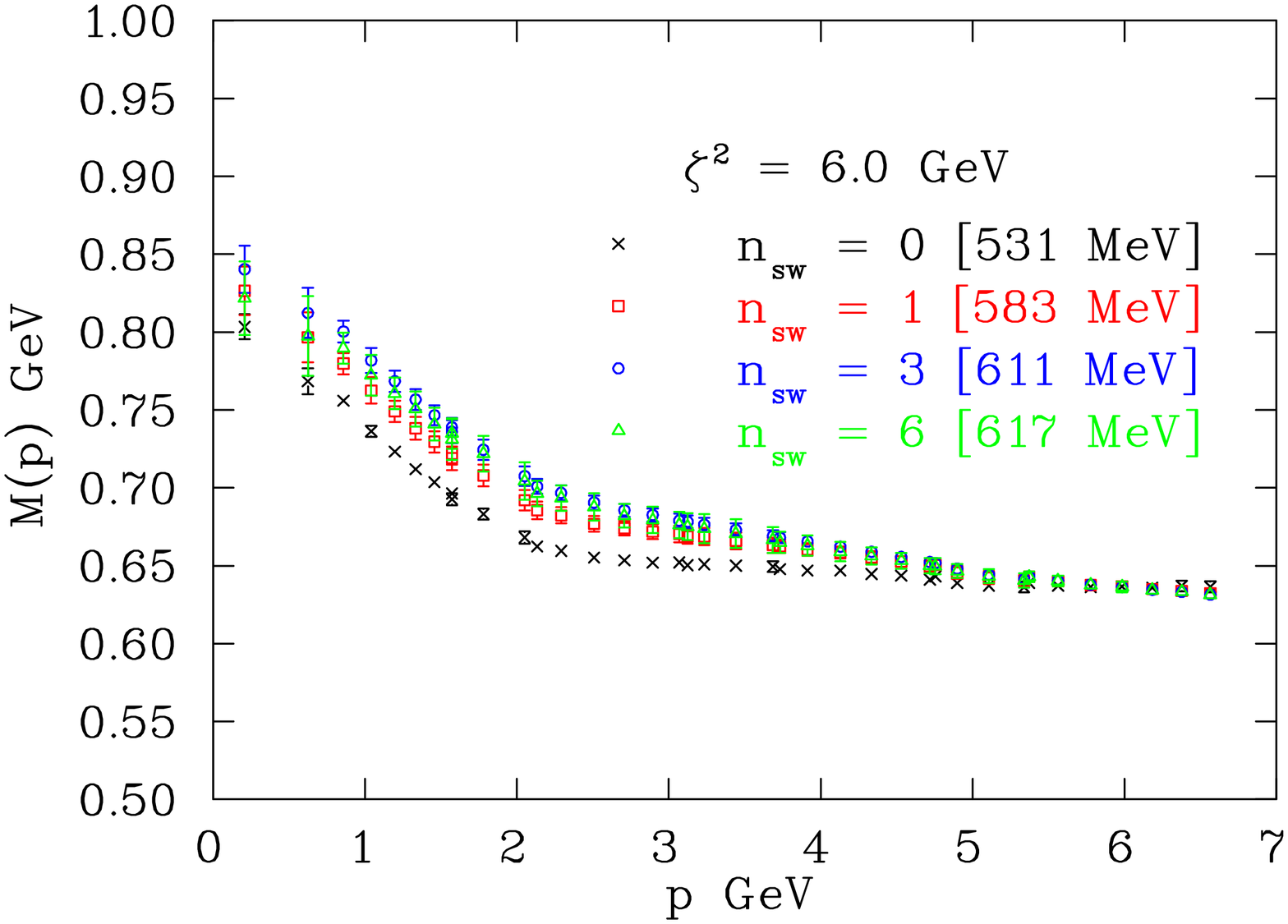} & 
\includegraphics[angle=90,width=0.45\textwidth]{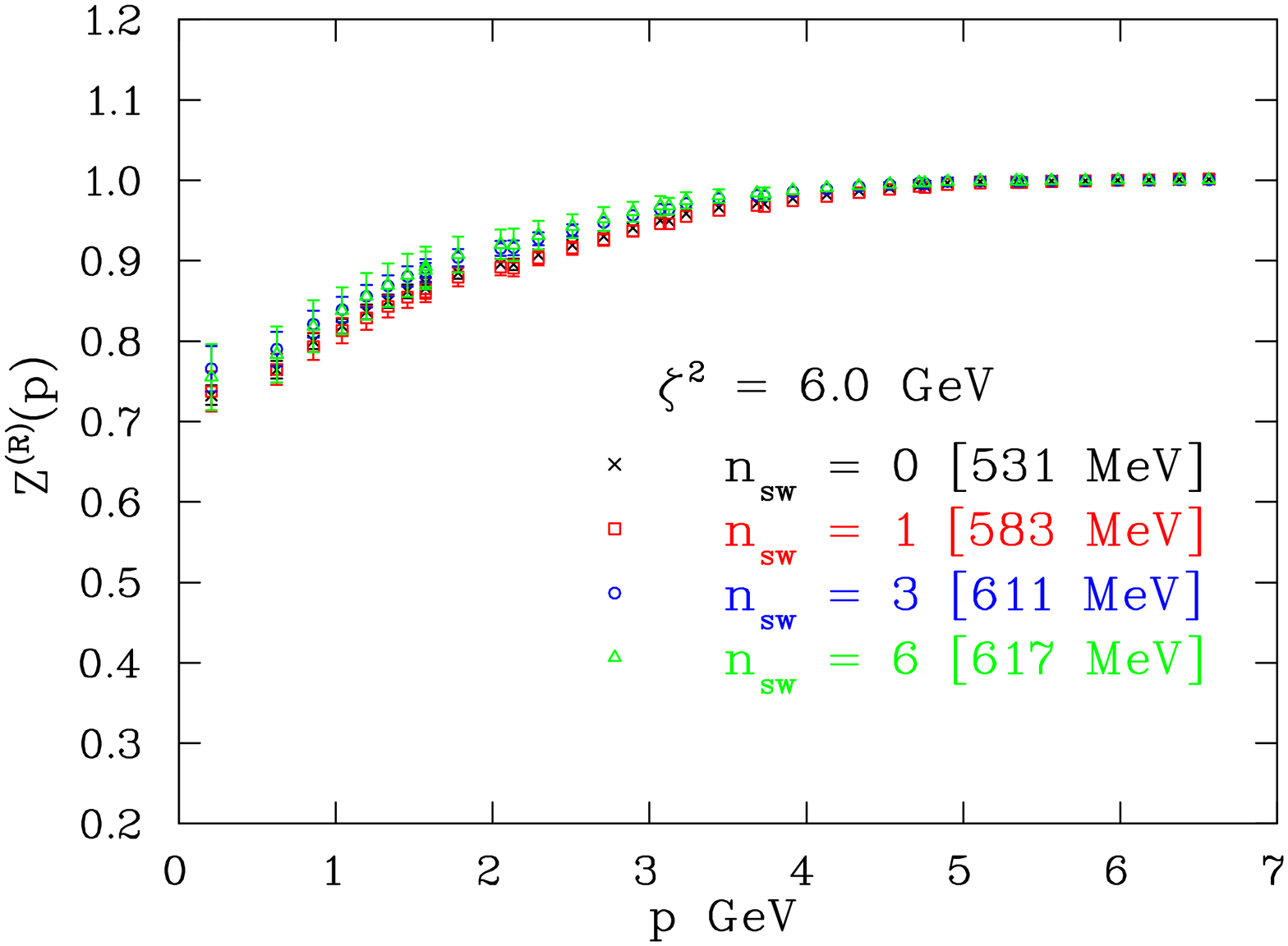}
\end{tabular}
\caption{The interpolated mass $M(p)$ and renormalization $\ZR$
  functions for the heavy bare quark mass, $m^0 = 531$~MeV, for the
  three choices of $\zeta$. The effective
  bare quark masses are given in square brackets. We see that for this
  value of $m^0$, the choices $\zeta = 3.9$, and $6.0$~GeV lead to
  large differences in the moderate and infrared momentum regions of
  $M(p)$. This indicates that the physics above approximately $3$~GeV
  has been spoiled by the smearing algorithm. In $\ZR$ we again find
  that the stout-link smearing algorithm introduces a small splitting
  in the infrared region.}
\label{0900res}
\end{figure*}

Finally we consider a larger choice of the bare quark mass, $m^0 =
531$~MeV. A consideration of the mass functions $M(p)$ given in
Fig.~\ref{0900res} reveals a strong dependence on the choice of
reference momentum $\zeta$. We see that a choice of either
$\zeta = 3.9$, or $6.0$~GeV leads to large discrepancies in both the
low and moderate momentum regions. With a choice of $\zeta =
2.0$~GeV we are able to obtain agreement in the low momentum region.

The dependence of $M(p)$ on $\zeta$ indicates that the suppression of
ultraviolet fluctuations by the smearing algorithm has spoiled the
physics of the theory above $\sim 2-3$~GeV, for this value of $m^0$.
These effects are clearly visible after just a single sweep of
smearing at this heavy bare quark mass. We further note that in the
case of $6$ sweeps and $\zeta = 6.0$~GeV, the mass function drops to
the bare quark mass.  This is a clear indication that the Compton
wavelength of the quark is small enough to reveal the void of
short-distance interactions following $6$ stout-link smearing sweeps.

\begin{figure}
  \includegraphics[angle=90,width=0.45\textwidth]{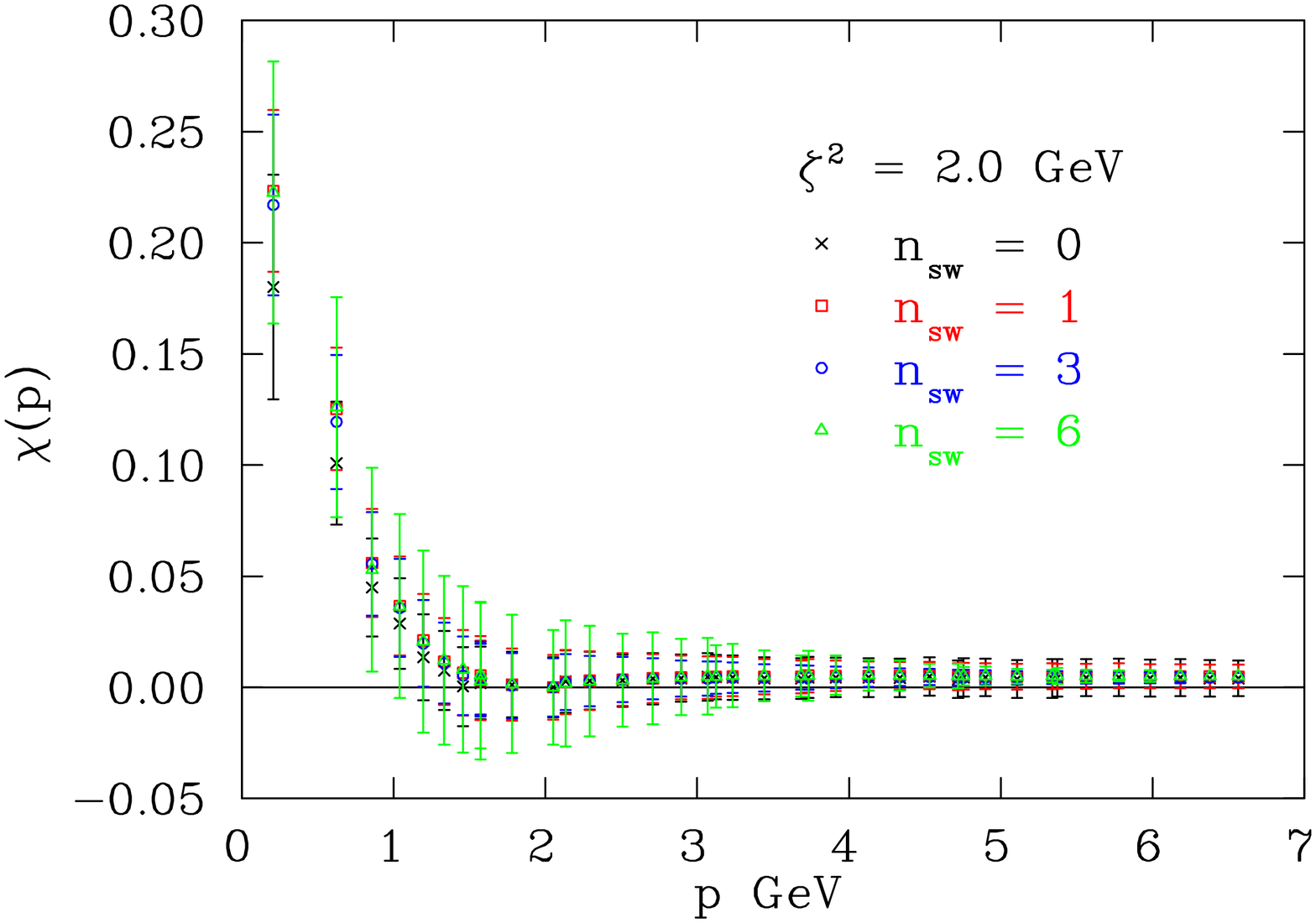}
  \caption{The difference $\chi(p) \equiv |
    Z^{(\rm{R})}_{\rm{light}}(p) - Z^{(\rm{R})}_{\rm{heavy}}(p) |$
    between the renormalization functions $\ZR$ for the heavy and
    small bare quark masses considered previously, with $\zeta =
    2.0$~GeV. We see that the difference rapidly approaches zero,
    indicating that the magnitude of the splitting introduced by the
    smearing algorithm is independent of the input bare quark
    mass. The differences at lower momenta are due to a flattening of
    $Z(p)$ as $m^0$ is increased.}
  \label{Zdiff}
\end{figure}

The renormalization functions $\ZR$ for a heavy bare quark mass of
$m^0 = 531$~MeV are also provided in Fig.~\ref{0900res}.  Apart from
the small splitting in the UV region, $\ZR$ still appears to be mostly
unaffected by the smearing algorithm.  In Fig.~\ref{Zdiff} we show the
differences in $\ZR$ between the smallest and largest bare quark
masses, where in order to examine the UV splitting we choose $\zeta
= 2.0$~GeV.  Figure~\ref{Zdiff} shows that the magnitude of the
splitting in $\ZR$ introduced by the smearing algorithm is unaffected
of the input bare quark mass.

The stout-link smearing procedure can save a large amount of compute
time in the calculation of hadronic physics. Not only is the Dirac
operator easier to invert but statistical errors are reduced
significantly.  The conclusion drawn from this study is that up to six
sweeps of stout-link smearing sweeps induces rather small effects on
the quark propagator for small and moderate bare quark masses, as
claimed by Durr, {et~al.}~\cite{Durr1, Durr2, Durr3}. After an
appropriate rescaling of the bare quark mass, the renormalized quark
propagator displays the same physics as the untouched configuration.
The only notable exceptions are order $2\%$ discrepancies in the
renormalization function for all quark masses and the most infrared
point of the lightest quark mass function. There an effect approaching
$2\sigma$ is revealed.

These subtle effects provide some evidence of a link between small
topologically nontrivial gauge field configurations linked to
dynamical chiral symmetry breaking through their production of
approximate zero-modes in the Dirac operator.  Upon smearing this
short distance physics is modified. 

Certainly the effects are subtle. However, they may require further
investigation in the event that fermion actions, in which all links of
the action are smeared, become the action of choice for calculating
the physics beyond the standard model.

\acknowledgments

We thank both eResearch SA and the NCI National Facility for generous
grants of supercomputer time which have enabled this project.  This
work is supported by the Australian Research Council. JBZ is partly
supported by Chinese NSFC-Grant No.  10675101.

\end{document}